# Integrative modelling of protein-glycan interactions with HADDOCK3.

Victor Reys, Marco Giulini, Alexandre M.J.J. Bonvin

Bijvoet Centre for Biomolecular Research, Department of Chemistry, Faculty of Science, Utrecht University, Padualaan 8, 3584 CH, Utrecht, the Netherlands.

## Abstract

Glycans are structurally diverse and flexible biomolecules that play key roles in many biological processes. Their conformational variability makes the modeling of their interactions with proteins particularly challenging.

This chapter presents a step-by-step protocol for modeling protein–glycan interactions using HADDOCK3, an integrative modeling platform that supports the inclusion of experimental or predicted interaction restraints and allows for flexible refinement of the solutions. The workflow is illustrated using the interaction between a linear homopolymer glycan, 4-β-glucopyranose, and the catalytic domain of the Humicola grisea Cel12A enzyme, for which an experimental X-ray structure is available as a reference. Detailed instructions are provided for input structure preparation, restraint definition, docking setup, execution, and result analysis.

Application of the protocol starting from unbound structures yields models of acceptable to medium quality, with interface–ligand RMSD values below 3Å. Although illustrated on a specific system, the protocol has been optimized and benchmarked on multiple protein–glycan complexes and is broadly applicable to similar systems, providing a framework for integrative modeling of protein–glycan interactions.

# 1. Introduction

A glycan is a molecule composed of different monosaccharide units, linked to each other by glycosidic bonds. Glycans are involved in a wide range of biological processes, such as cell-cell recognition, cell adhesion, and immune response. They are highly diverse and complex in their structure, as they can involve multiple branches and different linkages, namely different ways in which a glycosidic bond can connect two monosaccharides. The variety of monosaccharides, their number and the linkages between them makes the glycans a family of chemical compounds of great diversity. Their complexity, together with their flexibility makes the prediction of their three-dimensional (3D) conformation(s) a difficult task. While laborious, computational tools together with molecular dynamics simulations can now be used to tackle the effort of building accurate structural models of glycan in a matter of seconds [1,2], enabling to obtain in a straight forward manner structures for their use in downstream applications.

Not only do glycans belong to a fascinating family of biological compounds, they also have the beauty of interacting with high specificity to other biomolecular entities, such as proteins. The variety of potential glycans, together with their flexibility makes the prediction of their 3D conformation when interacting with a protein a challenging task. As structure prediction deep-learning tools may still struggle to provide accurate models for glycans and their interactions, due to a lack of co-evolution signals, hardly obtainable for glycans, molecular docking tools remain a useful choice when dealing with such entities. Among those, HADDOCK3 [3], an integrative modeling software, allows to define specific restraints to focus the sampling towards particular region of the space, and to introduce flexibility in both glycans and proteins, through short simulated annealing protocols based on molecular dynamics simulations, enabling to model (limited) binding-induced conformational changes that might be required to obtain better structural models of protein-glycan complexes.

In this chapter, we describe a detailed protocol for modelling protein-glycan interactions, using as an example the complex formed between a linear homopolymer glycan, 4-beta-glucopyranose and the catalytic domain of the *Humicola Grisea* Cel12A enzyme. The structure of this complex has been solved experimentally by X-ray crystallography (PDB ID: 1uu6 [4]), which allows us to evaluate the quality of the generated models. We provide comprehensive instructions for the use of HADDOCK3 in the context of integrative modelling of protein-glycan interactions. All the steps are described to perform a dedicated docking run with HADDOCK3, from obtaining and curating the input files, setting up a docking workflow and analysing the results. While this is illustrated with a specific protein-glycan interaction, this approach is generic and has been extensively benchmarked by Ranaudo et al., [5].

## 2. Materials

### 2.1 Software

HADDOCK3 [3] software package (freely available from https://github.com/haddocking/haddock3) and a functional Python3 (versions 3.9 to 3.14) installation on the operating system. See the installation instructions below.

### 2.2 Hardware

We have tested the installation on CentOS 7, RockyLinux, RedHat, and Apple M2 MacOS 13.5. A computer with multiple CPUs is preferable, to speed-up the computations that can be performed in parallel.

### 2.3 Data

Structural models of both protein and glycan in PDB format.

## 3. Methods

In this section, we outline the process to perform integrative modelling of protein-glycan interaction with HADDOCK3. Additionally, we offer instructions on how to extract residues potentially involved in the binding on the protein side and how to generate ambiguous interaction restraints from them.

To perform the integrative modelling of protein-glycan with HADDOCK3, the following steps should be followed:

1. Installation of HADDOCK3
2. Preparation and preprocessing of the input data, consisting of the initial structures and the distance restraint file used to guide the docking
3. Definition of a dedicated HADDOCK3 docking workflow specialised for protein-glycan interaction
4. Analysis of the results

In the following, all those steps are described in detail.

## 3.1 Installation of HADDOCK3

HADDOCK3 can be installed from the python package index (PyPI), or from its source code directly. Installation instructions for both cases are provided below. For this to work you must have a working Python3 (versions 3.9 to 3.14 supported) installation.

If you need to install Python3 visit: https://python.org and follow the installation instructions.

### 3.1.1 Installation of HADDOCK3 via pip

1. We recommend using Python's native virtual environment to manage the dependencies and avoid conflicts in your system (see note 1).

   ```
   python3 -m venv ~/.venv-haddock3
   ```

   ```
   source ~/.venv-haddock3/bin/activate
   ```

2. Install HADDOCK3 from the Python3 Package Index (https://pypi.org/project/haddock3/). For the long-term reproducibility of this protocol, we suggest to use the HADDOCK3 version 2026.3.0, as this is the version described here (see note 2):

   ```
   pip install haddock3==2026.3.0
   ```

### 3.1.2 Installation of HADDOCK3 latest version

1. We recommend using Python's native virtual environment to manage the dependencies and avoid conflicts in your system (see note 1).

   ```
   python3 -m venv ~/.venv-haddock3
   source ~/.venv-haddock3/bin/activate
   ```

2. Download the HADDOCK3 source code from its GitHub repository

   ```
   git clone https://github.com/haddocking/haddock3
   ```

3. Finally, you can install HADDOCK3 directly from its source code

   ```
   cd haddock3
   ```

   ```
   pip install .
   ```

4. Alternatively, you can install the optional packages required for HADDOCK3 code development, local documentation and MPI libraries by running the following command

   ```
   pip install ".[dev,docs,mpi]"
   ```

### 3.1.3 Testing your installation

By following those steps, you should have successfully installed HADDOCK3. To be able to use HADDOCK3 from other terminals, you will have to activate the virtual environment again with:

```
source ~/.venv-haddock3/bin/activate
```

You can test your installation by giving the following command:

```
haddock3 --help
```

As output you should see:

```
usage: haddock3 [-h] [--restart RESTART] [--extend-run EXTEND_RUN] [--setup]
[--log-level {DEBUG,INFO,WARNING,ERROR,CRITICAL}] [-v] workflow

positional arguments:
  workflow            The input configuration file path describing the
workflow to be performed

options:
  -h, --help          show this help message and exit
  --restart RESTART   Restart the run from a given step. Previous folders
from the selected step onward will be deleted.
  --extend-run EXTEND_RUN
                      Start a run from a run directory previously prepared
with the `haddock3-copy` CLI. Provide the run directory created
                      with `haddock3-copy` CLI.
  --setup             Only setup the run, do not execute
  --log-level {DEBUG,INFO,WARNING,ERROR,CRITICAL}
  -v, --version       show version
```

## 3.2 Preparing the input structures.

In order to follow this protocol, we recommend to create a specific directory for it where you will save all required files. E.g:

```
mkdir protein-glycan
cd protein-glycan
```

### 3.2.1 Preparing the protein

An unbound structure of the catalytic domain of the Humicola Grisea Cel12A enzyme is available on the Protein Data Bank [6], with PDBid 1OLR [7] (https://www.ebi.ac.uk/pdbe/entry/pdb/1olr), which is a high quality crystal structure solved at 1.2Å resolution, and therefore suitable for this docking experiment. Before its use in HADDOCK, this structure must be downloaded and curated to remove water molecules, potential small molecules from the crystallisation buffer or co-factors (if not involved in the binding), to keep only the spatial coordinates of the protein atoms and to attribute it a unique chain identifier. All those steps can be performed using the pdb-tools [8] python package, which is installed together with HADDOCK3.

1. First the structure must be downloaded from the Protein Data Bank

   ```
   pdb_fetch 1OLR > 1OLR.pdb
   ```

2. Then all heteroatoms are removed, to only keep amino-acids information

   ```
   pdb_delhetatm 1OLR.pdb > 1OLR_noHETATM.pdb
   ```

3. To remove all non-coordinate records from the file, run the following command

   ```
   pdb_keepcoord 1OLR_noHETATM.pdb > 1OLR_ATOM.pdb
   ```

4. Also, the chain ID of this structure must be set to A

   ```
   pdb_chain -A 1OLR_ATOM.pdb > 1OLR_ATOM_A.pdb
   ```

5. Finally, make sure that the obtained PDB file will be valid

   ```
   pdb_tidy -strict 1OLR_ATOM_A.pdb > 1OLR_ready.pdb
   ```

Alternatively, all those separate commands can be performed in a single command, by piping the output of one prior command to the next one, as follows (see note 3):

```
pdb_fetch 1OLR | pdb_tidy -strict | pdb_delhetatm | pdb_keepcoord | pdb_chain -A | pdb_tidy -strict > 1OLR_ready.pdb
```

### 3.2.2 Preparing the glycan

The unbound structure of the linear polymer composed of 4 beta-D-glucopyranoses linked by beta-1,4-glycosidic bonds will be modelled using the GLYCAM webserver [1]. This online tool allows to build the 3D structure of glycans that may not be present in the PDB.

In the following section, we will go step by step from building the desired glycan using the GLYCAM webserver to post-processing it to make it compatible with its use in HADDOCK.

1. First, go to the GLYCAM - Carbohydrate Builder online tool, accessible at https://glycam.org/cb/.

2. In "Step 1: Set Glycan Sequence", click on the "Glc" monosaccharide (the small icon with a blue circle). By doing so, the text displayed in "Sequence" should be updated to "`DGlcpb1-OH`".

3. To start building a di-saccharide, click again on the "Glc" monosaccharide. By doing so, the text displayed in Sequence should be updated to "`DGlcpb1-(...)DGlcpb1-OH`". Note the presence of the parenthesis in the text. This indicates that the linkage is not yet defined, and therefore must be specified to beta-1,4.

4. In the "Linkage" section under, from the β linkages choose "1-4". Once the selection is made, the Sequence text should be updated to `DGlcpb1-4DGlcpb1-OH` and the Linkage section made no more available. Other parameters do not need to be modified, as a beta-D-glucopyranose is desired and "D" is already the default parameter.

5. As you may have understood it already, steps 3 and 4 must be repeated two more times, to add the third and fourth monosaccharides to the sequence and obtain the desired 4 beta-D-glucopyratoses linked by beta-1,4-glycosidic bonds.

6. The final text displayed in the sequence should now be: "`DGlcpb1-4DGlcpb1-4DGlcpb1-4DGlcpb1-OH`"

7. Click on the "Done" button, which will trigger the modelling of the desired glycan. You should be quickly redirected to the "Step 3: Download" section of the Carbohydrate Builder tool.

8. From this page, on the table on the right, click on the PDB button located under the "Download Minimized Structure", initializing the download of the energy-minimised structure of the glycan (see note 4).

You should now have downloaded a file named "`DGlcpb1-4DGlcpb1-4DGlcpb1-4DGlcpb1-OH_structure_min-gas.pdb`", containing the 3D coordinates of the desired 4 beta-D-glucopyranoses linked by beta-1,4-glycosidic bonds. Make sure to move it into the directory you created for this tutorial.

Unfortunately, the structure cannot yet be used directly into HADDOCK, as the file contains issues that can be solved by a few post-processing steps using pdb-tools. Indeed, D-glucopyranose residue names do not follow the PDB convention which is also the one used in HADDOCK, and therefore must be modified to BGC. In addition, the residue indices have to be modified from 1 to 4 only. Finally, no chains are defined for this molecule, which is a requirement for HADDOCK3. Here are the required steps to manipulate the structure and make it HADDOCK compatible.

1. First, the aglycon (the -OH group) will be treated. For this the residue name ROH must be selected from the GLYCAM structure, its name replaced to BGC

   ```
   pdb_selresname -ROH DGlcpb1-4DGlcpb1-4DGlcpb1-4DGlcpb1-
   OH_structure_min-gas.pdb | pdb_rplresname -ROH:BGC | pdb_chain -B
   > aglycon.pdb
   ```

2. As a second step, the sugar part of the structure (residues 2 to 5) must be selected, residue names 0GB and 4GB must both be replaced to BGC and residue numbering should start from 1. All of these manipulations can be performed using the following command

   ```
   pdb_selres -2:5 DGlcpb1-4DGlcpb1-4DGlcpb1-4DGlcpb1-
   OH_structure_min-gas.pdb | pdb_rplresname -0GB:BGC |
   pdb_rplresname -4GB:BGC | pdb_reres -1 > sugar.pdb
   ```

3. Then, the two parts (aglycon and sugar) can be merged together to form our processed glycan

   ```
   pdb_merge aglycon.pdb sugar.pdb > glycan.pdb
   ```

4. Finally, the chain and segment identifier B will be set to the glycan, and the structure terminated with a TER statement

   ```
   pdb_chain -B glycan.pdb | pdb_chainxseg | pdb_tidy > glycan-B.pdb
   ```

By following those sets of commands, you should now have a valid glycan structure to be used as docking input in HADDOCK3, as it follows the appropriate BGC residue naming (see note 5), has a different chain identifier (B), is composed of 4 residues and terminated with a TER statement.

### 3.2.2 Downloading the reference structure

To evaluate the quality of the docking results, generated models can be compared to a reference structure if known. In this particular case, the complex has already been experimentally solved by X-ray crystallography (PDBid 1uu6 [4]). One can download this structure and use it directly in HADDOCK3 within the `caprieval` module, which is made to compare generated models against a reference structure using the CAPRI criteria. As the structure contains other ligands as well as water, the structure must be pre-processed prior to its use in a similar manner as done previously..

```
pdb_fetch 1UU6 | pdb_delresname -HOH,PG4 | pdb_tidy >
1UU6_target.pdb
```

By following those sets of command, the structure stored under the PDBid 1UU6 is first downloaded (`pdb_fetch`), then undesired solvent molecules, such as water (HOH) and tetraethylene glycol (PG4), are removed from the file (`pdb_delresname`), and finally the file is post-processed (`pdb_tidy`) to make sure it respect the appropriate formatting and saved under the file named `1UU6_target.pdb`.

## 3.3 Defining distance restraints to guide the docking

### 3.3.1 Identifying interacting residues

Incorporating experimental or predicted information in the modeling process is at the core of the HADDOCK software, which uses distance restraints to focus the search during the docking process toward models that maximize the consistency with the input data. Distance restraints, often represented as ambiguous interaction restraints (AIRs), must be defined between the two partners. In this docking scenario, we have information about the glycan binding site on the protein, but no knowledge about which monosaccharide units are relevant for the binding. In this particular case, all the four beta-D-glucopyranose units are at the interface, although this might not be true in general, especially when longer glycans are considered.

For the sake of this tutorial, the set of residues composing the binding site and interacting with the glycan can be taken from the reference crystal structure.
Here is the set of binding site residues within a 3.9Å distance cutoff from the glycan:

```
22,24,59,64,97,103,105,115,120,122,124,131,132,133,134,155,158,205,207
```

And at 5.0Å distance cutoff from the glycan:

```
9,22,24,59,61,63,64,66,97,99,103,105,114,115,120,122,124,130,131,132,133,134,155,158,160,205,207
```

Alternatively, when information about the binding site of the glycan is not available, residues potentially interacting with the glycan can be predicted using bioinformatics tools, such as PeSTo-Carbs [9]. This tool will predict the probability of residues to be interacting with a carbohydrate. Here are the instructions to predict the set of residues using the PeSTo-Carbs webserver (see note 6), which requires inputting a protein structure and selecting the chain of interest.

1. First, go to the PeSTo webserver [9] at https://pesto.epfl.ch/

2. Then, click on the "Upload PDB" green button, to load the protein structure we prepared above (`1OLR_ready.pdb`).

3. On the right of it, modify the prediction tool from "PeSTo" to "PeSTo-Carbs general"

4. At the bottom of the page, click on "Detect chains" to trigger chain detection from the input file

5. In this particular case, only one chain is present in your file, therefore the appropriate chain should be already selected by default

6. Finally, click on "submit" to launch the predictions.

7. Within a few seconds, the results page should open-up. This page is composed of two parts. The top one displays the structure with predicted interacting residues colored from white to red (white being low confidence interacting residues while red is used for high confidence ones). The bottom part is tabulating the same information, with the first three columns identifying residues (chain, residue names and residue index), while the last one contains the probability for a given residue to be in contact with a glycan. Note that only residues with probability above 0.5 are shown.

8. In our case, we want to gather all residue identifiers predicted with high confidence (e.g.: with a probability above 0.7). Note however that PeSTo renumbers the residues starting from 1, which is a shift of -1 in this particular case. We therefore have to add +1 to the residue numbers predicted by PeSTo-Carbs. The corresponding coma separated list is:

```
9,20,22,24,61,64,66,103,105,114,115,120,122,134,142,147,155,158,199,201,205
```

You can note how similar the two sets of residues identified by the two different methods are. One was directly extracted from the reference file while the other was predicted by PeSTo-Carbs software. Be careful, as there is a strong bias in this particular case, as the target complex pdb structure 1UU6 was part of the training set of PeSTo-Carbs, making the prediction quite precise in this case.

### 3.3.2 Generating the ambiguous interaction restraints file

In this section we will define the restraints that will guide the docking of the protein and glycan structures. For this, we need to define two sets of interacting residues, one for each input molecule. For the protein, we will only define active residues (based on the identified binding site above), while for the glycan we define all four residues as passive (passive meaning that if a residue is not at the interface in the resulting model it will not generate an energetic penalty). In the following section, we introduce the use of the haddock-restraints online web server for the generation of such restraints.

1. First, go to the haddock-restraints online web server at https://wenmr.science.uu.nl/haddock-restraints, where you should be landing in the HADDOCK restraint generator tool

2. In the "Interactor 1" field, which will be used to define binding site residues of the protein, fill up the "Active Residues" box with the comma separated list of residues extracted from the reference structure with a 3.9Å contact cutoff:

   `22,24,59,64,97,103,105,115,120,122,124,131,132,133,134,155,158,205,207`

3. Then, in the same "Interactor 1" field, provide as reference structure the PDB file of the unbound protein we have just prepared. For this click on Select File and select the `1OLR_ready.pdb` file. By providing a 3D structure we will be able filter out residues that are not solvent accessible.

4. Click on "See advanced options" and then turn on the "Filter Buried" option and set the "Buried cutoff" to 0.15, which means that residues that have a relative solvent accessibility below 15% will be filtered out from the restraints

5. Then, in the "Target interactor", select 2, which will be the glycan.

6. All other parameters can be left by default, as the selected chain is by default A, which is the chain ID that has been defined for the protein.

7. In the "Interactor 2" field, which will be used to define interacting residues of the glycan, fill up the "Passive Residues" with the comma separated list of the four residues that are composing the glycan.

   `1,2,3,4`

8. Then, in the "Target interactor", select 1, which will be the protein.

9. To generate the restraints, click on the "Generate restraints" button, at the bottom of the page. This will trigger the generation of several distance restraints, each starting with the "assign" statements, that are understood by CNS (Crystallography and NMR System [10], the computational engine of HADDOCK) and used to guide the docking (see note 7).

10. Finally, click on the "download" button to download the content of the file on your computer. Move that file to the location where you have your prepared PDB files.

## 3.4 Setting up a docking workflow specialised for protein-glycan interaction and executing it with HADDOCK3

Now that all preliminary steps related to the obtention and generation of input files have been completed (two PDB files and the restraints), this section will focus on the definition of a HADDOCK3 protein-glycan specific workflow. Indeed, in HADDOCK3, workflows are flexible. A user must define and order the modules relevant for the particular scenario, together with any parameter that needs to be modified and restraint files that will be used in a given module. This flexibility in the desired workflow is new to HADDOCK3 and its potential is fully harvested in this protein-glycan docking scenario. Compared to the HADDOCK2.X series (and its webserver [11]), it allows to add a clustering step after the rigid-body docking, enabling to forward up to 20 models from each cluster, independently of their scores, to the flexible refinement stage, which is something that was not possible in the previous versions. In the following section we will describe how to generate a valid docking workflow used by HADDOCK3 and how to launch the tool itself.

The HADDOCK3 code comes with a series of examples for different docking, refinement and scoring scenarios that can serve as a basis to define a new workflow. See for that: https://github.com/haddocking/haddock3/tree/main/examples (see note 8)

### 3.4.1 Preparing the docking workflow

The HADDOCK3 configuration file follows a TOML-like format, with some slight adjustments. It is composed of two main sections; the general parameters and the list of consecutive modules to be executed sequentially.

In the general parameters, only two of them are mandatory, and for the remaining ones the default values will be used.

1. `run_dir` : Defines the name of the run directory where the results will be written.
2. `molecules` : Defines the list of input molecules, which in our case are the protein receptor (`1OLR_ready.pdb`) and the glycan (`glycan-B.pdb`).

Once the default parameters are defined, the list of modules must be added, in the appropriate order. Each module name is encompassed by square brackets. Each module has its own set of parameters that can be modified by the user to optimize their values for a specific workflow (see note 9). Without specifying a parameter value, the default one will be used. In the following, we will describe the ordered list of modules and the parameters that must be modified or defined (e.g. the restraint files), when needed. *Note* that the workflow described here is adapted from the `docking-protein-glycan-ilrmsd-full.cfg` example provided with the haddock3 code.

1. `topoaa` : Generate the all-atom topology and parameters for each input molecule and build any missing atoms in the input structures. This is a requirement as they are used by the CNS engine. This module is always defined as the first module to ensure the usability of downstream CNS modules (in this case; rigidbody and flexref).

2. `rigidbody` : Performs the rigid-body docking between the two partners (by default 1000 models will be generated).

a. `ambig_fname` : Defines the path to the ambiguous interaction restraints (names "restraints.tbl") guiding the docking during the integrative modelling process.

b. `w_vdw` : Defines the weight of the Van der Waals component of the HADDOCK scoring function. Note that tuning the value of this parameter to 1.0 (instead of the default value of 0.1) increases the Van der Waals energetic term contribution to the scoring function, which would otherwise be too influenced by the non-specific electrostatic interactions between the multiple hydroxyl groups present on the glycan (see note 10).

3. `caprieval` : Compare the models against a reference structure using the CAPRI metrics (see note 11).

    a. `reference_fname` : Defines the path to the reference file. In our case, the experimental structure of the complex is available (PDBid 1uu6, pre-processed under the name `1UU6_target.pdb`) and therefore can be used as reference to evaluate the quality of the generated complexes (see note 12).

    b. `keep_hetatm` : This parameter is set to true, allowing to keep HETATM records in the reference file (the glycan is defined as HETATM).

4. `ilrmsdmatrix` : Compute the interface-ligand root-mean-squared deviation (RMSD) matrix. This fits the models onto the backbone of the interface residues of the protein and then calculates the RMSD on the ligand, the glycan (see note 13).

5. `clustrmsd` : Perform the clustering of the previously computed RMSD matrix

    a. `criterion` : Defines how the clustering must be performed. In this case, the "`maxclust`" criterion will be used, allowing to define a given number of desired clusters.

    b. `n_clusters` : Defines the number of desired clusters. In this case we want to generate 50 most dissimilar clusters.

6. `seletopclusts` : Performs the selection of best scoring models among best scoring clusters

    a. `top_clusters` : Number of best scoring clusters from which models should be forwarded to the next stages. In this case we want to gather all 50 clusters.

    b. `top_models` : Number of best scoring models to take from each selected cluster. In this case, we want to take up-to 20 models per cluster.

7. `caprieval` : Compare the selected models against a reference structure using the CAPRI metrics.

8. `flexref` : Perform the flexible refinement of residues at the interface. This is a simulated annealing protocol based on torsional angle space molecular dynamics. By default, interface residues are treated as flexible in the last two stages of the protocol, first side-chains and then both backbone and side-chains.

   a. `ambig_fname` : Defines the path to the ambiguous interaction restraints guiding the docking during the integrative modelling process.

9. `caprieval` : Compare the models against a reference structure using the CAPRI metrics.

10. `ilrmsdmatrix` : Compute the interface-ligand RMSD matrix of the current set of complexes

11. `clustrmsd` : Perform the clustering of the previously computed RMSD matrix

    a. `criterion` : Defines how the clustering must be performed. In this second case, the "`distance`" criterion will be used, allowing to define a distance threshold used to separate clusters.

    b. `clust_cutoff` : Defines the distance cutoff used to group complexes together. In this case the value is set to 2.5 Ångstroms.

    c. `linkage` : Defines the type of linkage used to generate the dendrogram, where the "`average`" method is used.

12. `seletopclusts` : Performs the selection of best scoring models among best scoring clusters

    a. `top_models` : Number of best scoring models to take from each selected cluster. In this last stage, only the top 4 models from each cluster will be forwarded to the next step.

13. `caprieval` : Final comparison of the selected complexes against the reference.

Below is the content of the resulting HADDOCK3 workflow, which was described in plain word above, with the desired parameters values for each module:

```
# ====================================================================
# protein-glycan docking using information about the protein binding site
# and no information on the glycan side
# ====================================================================
# ==========================
# General parameters section
# ==========================
run_dir = "run_prot_glyc"
molecules = [
```

```
    "1OLR_ready.pdb",
    "1UU6_l_u.pdb",
]

# ==============================================
# List of modules to be performed consecutively
# ==============================================

[topoaa]

[rigidbody]
ambig_fname = "restraints.tbl"
w_vdw = 1

[caprieval]
reference_fname = "1UU6_target.pdb"
keep_hetatm = true

[ilrmsdmatrix]

[clustrmsd]
criterion = "maxclust"
n_clusters = 50

[seletopclusts]
top_clusters = 50
top_models = 20

[caprieval]
reference_fname = "1UU6_target.pdb"
keep_hetatm = true

[flexref]
ambig_fname = "restraints.tbl"

[caprieval]
reference_fname = "1UU6_target.pdb"
keep_hetatm = true

[ilrmsdmatrix]

[clustrmsd]
criterion = "distance"
clust_cutoff = 2.5  # The clustering distance is set top 2.5 Angstroms
linkage = "average"

[seletopclusts]
top_models = 4

[caprieval]
reference_fname = "1UU6_target.pdb"
keep_hetatm = true

# ==============================================
```

The content of the workflow must be written in an ASCII text file that is used by HADDOCK3. For the sake of this protocol, we will call it "`protein-glycan-workflow.cfg`".

### 3.3.2 Running the workflow

All input files are now generated and ready to be used in HADDOCK3. In the same directory, you should have:
- three different PDB files containing:
    - the structure of the protein ("`1OLR_ready.pdb`")
    - the structure of the glycan ("`glycan-B.pdb`")
    - the reference structure ("`1UU6_target.pdb`")
- the ambiguous distance restraints file ("`restraints.tbl`")
- the HADDOCK3 workflow configuration file ("`protein-glycan-workflow.cfg`")

With all these files at hand, and the HADDOCK3 environment activated, typing the `haddock3` command followed by the path to the HADDOCK3 configuration file, will launch your docking workflow.

```
haddock3 protein-glycan-workflow.cfg
```

*Note* that if you want to avoid HADDOCK3 using all the CPU cores of your system, you can limit their number by tuning, in the general parameter section defined at the beginning of the workflow, the parameter defining the maximum number of cores to use, e.g.: `ncores = 10`

With all these files, and the HADDOCK3 environment activated, the `haddock3` command will launch your docking workflow. As an indication of run time, this workflow completed in less than 31 minutes using 10 cores (`ncores = 10`) on a M2 MacBook Pro and about 12 1/2 minutes on a Linux server with AMD EPYC 7502 32-Core Processors using 50 cores.

## 3.4 Analysing the results

### 3.4.1 Navigating through the results of the docking run

HADDOCK3 will start by validating the workflow file, making sure the parameters and their values fall within the accepted range and input files can be found in the system. The run directory is created based on its name given in the configuration file and input files are then stored in the

`data` directory. Then, HADDOCK3 will sequentially process each module, creating a sequentially numbered directory for each of them and storing the generated files in it. Those generated files can be in the form of PDB structures (for sampling, refinement or selection modules), figures stored as HTML files (mainly for analysis modules that generated plots) or csv files (mostly for scoring modules). At the end of the run, two additional steps are performed by default: an analysis step (in the `analysis` directory), generating tables and figures of model quality for each `caprieval` step and a tracing-back step (in the `traceback` directory), allowing to traceback a given complex throughout the various stages of the workflow.

All the generated directories and files can be listed using the `ls` command:

```
ls run_prot_glyc/
```

which should give as output:

```
00_topoaa
01_rigidbody
02_caprieval
03_ilrmsdmatrix
04_clustrmsd
05_seletopclusts
06_caprieval
07_flexref
08_caprieval
09_ilrmsdmatrix
10_clustrmsd
11_seletopclusts
12_caprieval
analysis
data
log
traceback
```

For example, the `11_seletopclusts` directory contains the selected models from each cluster. The clusters in that directory are numbered based on their rank, i.e. `cluster_1` refers to the top-ranked cluster. Information about the origin of these files can be found in that directory in the `seletopclusts.txt` file.

### 3.4.2 Evaluating the quality of the generated models

The simplest way to extract ranking information and the corresponding HADDOCK scores is to look at the `X_caprieval` directories (which is why it is a good idea to have it as the final module, and possibly as intermediate steps, even when no reference structures are known). This directory will always contain a `capri_ss.tsv` file (*ss* standing for Single Structure), which contains the model names, rankings and statistics, such as the HADDOCK score (score), interface-RMSD (irmsd), Fraction of Native contacts (fnat), ligand-RMSD (lrmsd), interface-ligand-RMSD (ilrmsd), DockQ [12] score (dockq), and global RMSD (rmsd) for each analysed models. The iRMSD, lRMSD and Fnat metrics are the ones used in the blind protein-protein prediction experiment CAPRI [13] (Critical PRediction of Interactions).

Following the protein-protein complexes CAPRI criteria, the quality of a model is defined as:

- **acceptable model**: i-RMSD < 4Å or l-RMSD<10Å and Fnat > 0.1
- **medium quality model**: i-RMSD < 2Å or l-RMSD<5Å and Fnat > 0.3
- **high quality model**: i-RMSD < 1Å or l-RMSD<1Å and Fnat > 0.5

But these metrics are defined for protein-protein complexes. As glycans are typically smaller, it is best to use stricter metrics to assess the quality of the models. In the case of information-driven protein-glycan docking, the Fnat term is less relevant, as most contacts will typically be satisfied. Instead, for protein-glycan modelling, focusing on the interface-ligand RMSD (ilRMSD) has been recently proposed by A. Ranaudo *et al.,*[5] where the quality of the complex has been redefined as:

- **near acceptable model**: il-RMSD < 4Å
- **acceptable model**: il-RMSD < 3Å
- **medium quality model**: il-RMSD < 2Å
- **high quality model**: il-RMSD < 1Å

The ilRMSD is calculated by first superimposing the model onto the reference using the interface residues of the receptor (the protein in this case) and calculating the RMSD of the ligand (the glycans in this case). Depending on the length of the glycan, near-acceptable quality models can be considered decent for long linear glycans (like the one used), but for smaller glycans the quality of the models should be higher.

### 3.4.3 Visualising the scores and their components

Since multiple `caprieval` modules were placed at various stages of the workflow, between each sampling, selection and refinement modules, we can assess the quality of the generated models at the various stages of our workflow and evaluate if progress was made from one step to another. Particularly, the impact of the flexible refinement module (`flexref`) is worth the investigation, as potential improvements from the rigid-body docking can be observed.

This can be performed by comparing the content of the `capri_ss.tsv` files (*ss* standing for single structure), between step 6 and 8. The ilRMSD values are listed in column 8 of those files. To find the lowest ilRMSD value you can sort the file numerically based on column 8 and extract

the top model. This can be performed at the command line level with the following unix commands:

```
sort -nk8 run_prot_glyc/06_caprieval/capri_ss.tsv | head -2

sort -nk8 run_prot_glyc/08_caprieval/capri_ss.tsv | head -2
```

These commands are sorting the `capri_ss.tsv` files based on the 8th column and only displaying the first two lines (the header and the first line containing the best ilRMSD model). By comparing the values obtained after the rigidbody-stage and the flexible refinement one, you can observe a reduction of the lowest ilRMSD for the best model, improving from 3.5Å to 2.5Å, hence generating an acceptable quality model following the previously described criteria (see note 14). This shows that from a poor quality model, the flexible refinement, together with the restraints, allowed to apply sufficient conformational changes to improve the modelled complex. This is also reflected in the fraction of native contacts (Fnat) which improves from 0.59 to 0.77 (see note 15).

A more convenient way to visualise the results is by using the content present in the `analysis` directory, which contains post-processing analyses for each of the `caprieval` modules, where the data are presented as interactive tables and plots, all grouped in the `report.html` file. To visualise the quality of the models that reached the final step of the docking workflow, this file can be opened using your favourite web browser.

The report web-page is divided into three major sections (Figure 1). The first one holds all the metrics (Clusters details, HADDOCKscore and its components, and CAPRI metrics) related to each cluster (Figure 1A). By default, clusters are sorted by HADDOCKscore. To rapidly analyse what is the quality of the best models generated, the table columns can be sorted by clicking on the row header, for example on DockQ. By doing so, the table reorders by this metric, and readability increases. The second part contains various scatter plots which allow you to assess the convergence of the models. They show the HADDOCK scores and its components plotted against various evaluation metrics (Figure 1B), where each data point represents a model, coloured by its cluster ID. These are interactive plots made using *plotly* [14], meaning that by hovering with your mouse on a particular data point will make its statistics appear (e.g. model name and corresponding metrics), and selecting a particular part of the plot will be zooming towards it. As we provided the reference structure to the various `caprieval` modules, the plots allow us to directly assess the accuracy of the generated models. If no reference was provided, the best scoring model of each stage will be selected as reference, meaning that the plots will contain a point at RMSD=0 or Fnat=1 (see note 12). Finally, at the bottom are found several box plots, representing the distribution of scores for each cluster (Figure 1C).

# A  Analysis report of step 12_caprieval

| Cluster Rank → | 1 | 2 | 3 | 4 | 5 | 6 | 7 | 8 | 9 | 10 |
|---|---|---|---|---|---|---|---|---|---|---|
| Cluster ID | 1 | 22 | 11 | 2 | 13 | 10 | 7 | 27 | 9 | 8 |
| Cluster size | 4 | 4 | 4 | 4 | 4 | 4 | 4 | 4 | 4 | 4 |
| HADDOCK score [a.u.] | -159.99 ± 2.41 | -149.89 ± 8.2 | -147.33 ± 8.15 | -144.44 ± 11.05 | -142.06 ± 6.02 | -136.97 ± 4.59 | -134.97 ± 12.37 | -128.8 ± 5.45 | -127.88 ± 2.25 | -124.12 ± 1.8 |
| Van der Waals Energy | -47.05 ± 1.84 | -38.7 ± 9.34 | -43.23 ± 7.59 | -33.88 ± 5.34 | -33.7 ± 2.4 | -39.32 ± 3.99 | -41.16 ± 3.7 | -36.51 ± 5.78 | -43.32 ± 3.07 | -29.51 ± 2.41 |
| Electrostatic Energy | -99.65 ± 1.3 | -98.23 ± 12.85 | -91.26 ± 8.92 | -102.72 ± 15.38 | -102.22 ± 6.69 | -89.22 ± 6.76 | -81.75 ± 15.43 | -81.01 ± 9.87 | -71.51 ± 6.94 | -88.11 ± 5.1 |
| Restraints Energy | 19.43 ± 10.94 | 12.94 ± 9 | 23.69 ± 20.14 | 42.96 ± 21.44 | 64.69 ± 16.39 | 38.79 ± 20.78 | 32.37 ± 15.67 | 40.35 ± 11.52 | 22.36 ± 11.04 | 106.03 ± 21.94 |
| Desolvation Energy | -5.14 ± 0.64 | -4.19 ± 1.61 | -4.86 ± 2.2 | -2.14 ± 2.94 | -3.04 ± 1.21 | -2.83 ± 1.47 | -5.25 ± 1.39 | -6.07 ± 0.62 | -5.81 ± 1.7 | -8.52 ± 1.12 |
| interface RMSD [A] | 1.06 ± 0.06 | 4.21 ± 0.05 | 1.3 ± 0.05 | 4.11 ± 0.04 | 4.17 ± 0.03 | 4.43 ± 0.05 | 1.33 ± 0.1 | 1.61 ± 0.17 | 4.33 ± 0.02 | 4.33 ± 0.02 |
| ligand RMSD [A] | 2.79 ± 0.2 | 11.8 ± 0.14 | 3.52 ± 0.16 | 11.5 ± 0.1 | 11.65 ± 0.1 | 12.48 ± 0.17 | 3.69 ± 0.3 | 4.62 ± 0.51 | 12.11 ± 0.08 | 12.31 ± 0.05 |
| interface-ligand RMSD [A] | 2.79 ± 0.2 | 11.79 ± 0.14 | 3.52 ± 0.17 | 11.5 ± 0.1 | 11.66 ± 0.09 | 12.48 ± 0.16 | 3.69 ± 0.3 | 4.61 ± 0.51 | 12.12 ± 0.08 | 12.33 ± 0.05 |
| Fraction of Common Contacts | 0.73 ± 0.02 | 0.27 ± 0.02 | 0.74 ± 0.02 | 0.25 | 0.26 ± 0.01 | 0.33 ± 0.03 | 0.65 ± 0.03 | 0.49 ± 0.03 | 0.26 ± 0.01 | 0.27 ± 0.02 |
| DOCKQ | 0.77 ± 0.02 | 0.24 ± 0.01 | 0.72 ± 0.02 | 0.24 | 0.24 | 0.25 ± 0.01 | 0.69 ± 0.03 | 0.58 ± 0.04 | 0.23 | 0.23 ± 0.01 |
| Buried Surface Area [A^2] | 1008.48 ± 13.39 | 1005.69 ± 28.45 | 1034.36 ± 45.97 | 999.81 ± 36.84 | 957.64 ± 19.42 | 948.11 ± 25.39 | 1003.96 ± 19.79 | 924.93 ± 34.62 | 947.88 ± 15.32 | 858.48 ± 28.48 |
| Nr 1 best structure | ↓ Download ◉ View | ↓ Download ◉ View | ↓ Download ◉ View | ↓ Download ◉ View | ↓ Download ◉ View | ↓ Download ◉ View | ↓ Download ◉ View | ↓ Download ◉ View | ↓ Download ◉ View | ↓ Download ◉ View |

B
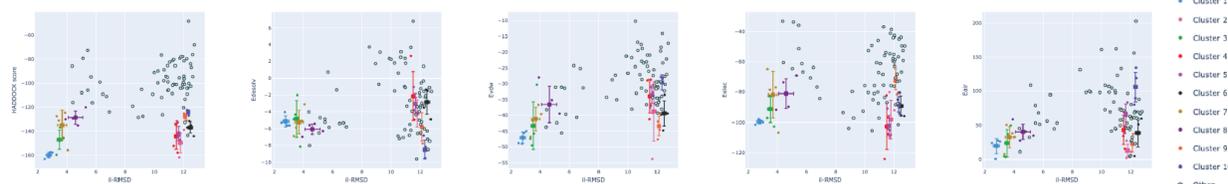

C
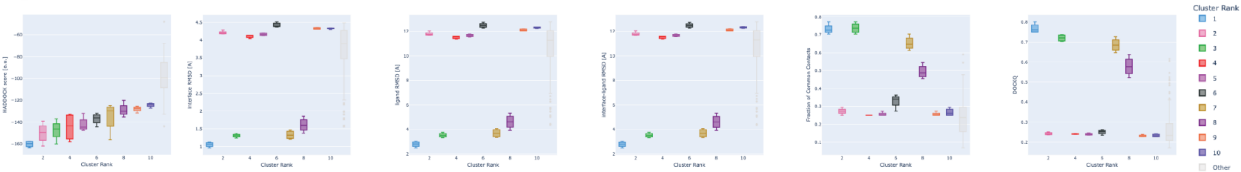

**FIGURE 1:** Visual representation of the content present in the analysis report for the step 12 `caprieval`. **A)** Picture of the interactive table, displaying the clusters information, scoring of their models and their components, and CAPRI quality metrics, for the top 10 best clusters, sorted by the HADDOCK score (which identifies the best cluster in terms of interface ligand RMSD on top in this particular case). **B)** A selection of scatter plots showing for each individual model the HADDOCK scores and its components vs the interface ligand RMSD to the reference target structure. **C)** A selection of box plots, displaying the distribution of scores of the various clusters. Note that only the top 10 clusters are color-coded in the B and C panels.

If your run has nicely converged to low HADDOCKscore solutions (the lower the score the better the model), you should see in the various RMSD/Fnat plots, a nice "binding funnel" toward the best model (or the reference if provided). If multiple solutions (clusters) have overlapping scores, you might see several such funnels, or in the worst case, no convergence at all and a wide distribution of points without a clear winner. In this particular case, this is indeed what we are observing (see Figure 1): The top 5 clusters are all overlapping in their scores considering their standard deviations. The cluster with the best model is cluster number 4, which actually also contains the overall best scoring model.

The "funnel" behavior can best be analysed by considering all models. For this we should analyse the rigidbody and flexref models with their corresponding analysis reports found under `analysis/02_caprieval` and `analysis/08_caprieval`, respectively. This is illustrated in Figure 2 where the convergence of the `rigidbody` (A) and `flexref` (B) stages is shown by plotting the HADDOCK score against the interface-ligand-RMSD. Pay attention to the difference in the scale of the scores between the rigidbody docking stage and the flexible refinement which leads to much better scores and more accentuated funnels (two of them with the slightly better scoring one identifying the correct solution). Finally, you can appreciate how the lowest scoring model at `flexref` (with -184 HADDOCKscore and 6.6Å ilRMSD with respect to the reference structure) is not kept after the clustering step `ilrmsdmatrix` and `clustrmsd` (C), as this model did not cluster with anything similar, thus allowing to discard rare conformations, potentially well evaluated with the scoring function, from converged conformations.

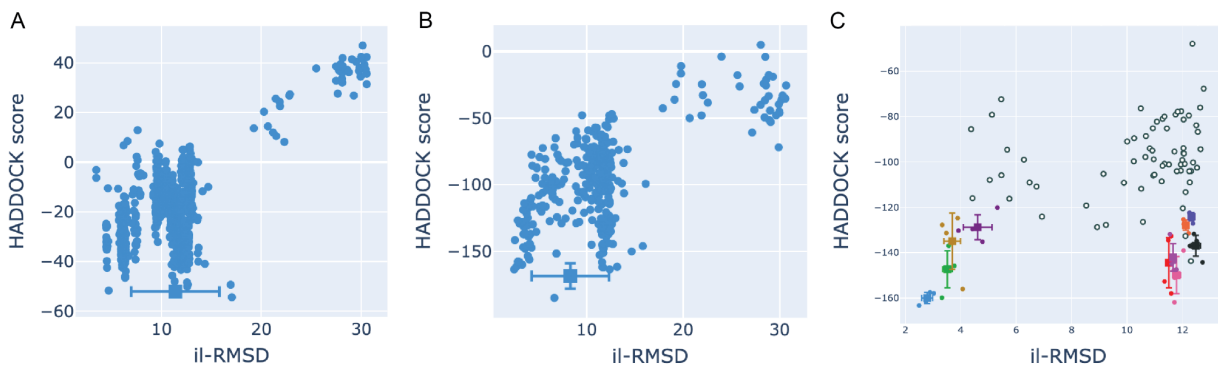

**FIGURE 2:** Scatter plots displaying individual models at two different stages of the workflow. **A)** After rigidbody docking (1000 models); several binding funnels can be observed. **B)** After flexible refinement, two major binding funnels (at ~3Å il-RMSD and ~12Å il-RMSD) are now present, displaying a convergence towards two distinct binding modes. **C)** After clustering and keeping only four models per cluster (the top 10 clusters are color-coded, the remaining clusters are all shown as white circles); singleton are discarded and main binding modes are kept, hence removing potential false positive hits from the pool of models. The squares and crosses are indicating the averages and standard deviations, respectively, calculated on the top four models (of all for A and B, and of each cluster in C). Note that these plots are automatically generated by HADDOCK3.

Finally, since we have in this particular tutorial the reference structure at hand, you can compare the best generated model (the best model of the best scoring cluster) with your favorite visualisation software. The overlay of the two structures is shown in Figure 3.

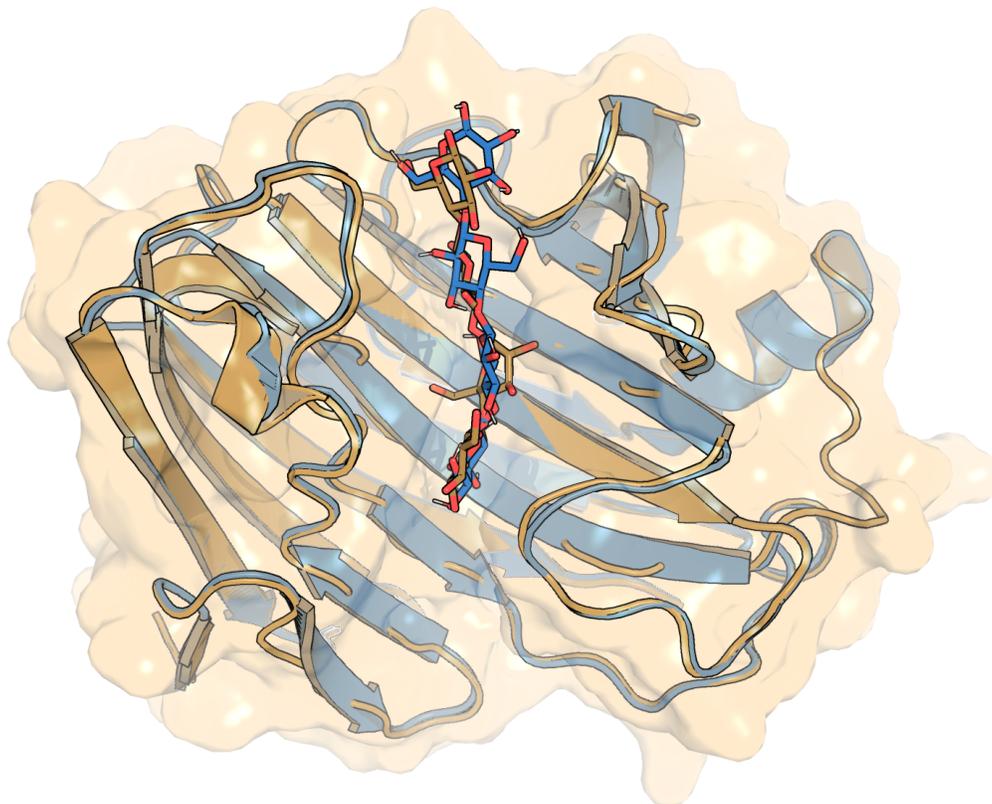

**FIGURE 3:** View of the best scoring model (blue) superimposed onto the reference crystal structure (1UU6) (wheat color). The interface-ligand-RMSD of this model from the reference is 2.49Å, falling into the acceptable model quality following the glycan-specific CAPRI criteria described above.

## 4. Conclusions

In this protocol, we have demonstrated the use of HADDOCK3 to predict the structure of a protein-glycan complex using predicted information about the protein binding site. We have shown how to prepare the PDB files for docking, define the ambiguous interaction restraints, and set up the docking protocol. We have also discussed the analysis of the docking results and the comparison with the reference structure.

For the selected system, the linear homopolymer glycan, 4-beta-glucopyranose binding to the catalytic domain of the *Humicola Grisea* Cel12A enzyme, starting from unbound conformation input structures, the proposed protocol resulted in models with interface-ligand RMSD under 3Å, falling into the acceptable quality range. The presented results also nicely illustrate that multiple clusters can be obtained with overlapping scores, while the best model does point to the right solution. Analysis of docking data remains therefore a delicate exercise and ideally additional data should be available to validate the models. Finally, while we focussed on one particular protein-glycan example, the illustrated protocol has been optimised and benchmarked on several protein-glycan complexes [5] and should be applicable to other similar systems. We refer to the original publication for more details on what performance can be expected depending on the length and complexity (branching) of the glycans.

## 5. Data availability

All input data and the results of the HADDOCK3 workflow described in this chapter are available from Zenodo (DOI: 10.5281/zenodo.18963538).

## 6. Acknowledgments

The authors acknowledge funding from the European High Performance Computing Joint Undertaking projects BioExcel (Center of Excellence for Computational Biomolecular Research) (101093290) and GANANA (Europe-India Partnership for Scientific High-Performance Computing) (101196247).

## 7. Notes

1. When HADDOCK3 is installed under a dedicated virtual environment, you will need to run the source command every time you start a new terminal to use it. Alternatively, the sourcing of the environment could be added directly in your `~/.bashrc` file.

2. You can also install the latest stable release of HADDOCK3 by simply not specifying any version while running the pip install command:

    ```
    pip install haddock3
    ```

3. Before blindly applying a set of pdb-tools commands to a new protein structure, it is always recommended to visualise the structure to check which chain(s) are relevant (removing irrelevant part of a structure - e.g. other chains/domains not involved in the binding - will speed up the computations) and which small molecules should be discarded or not (e.g. a co-factor like ATP might be important for the binding).

4. On the GLYCAM webserver, depending on the size of your browser window, the location of the table can be located under the displayed structure.

5. Not all glycans are supported in HADDOCK. The complete list of available glycans can be accessed at https://wenmr.science.uu.nl/haddock2.4/library.

6. If the PeSTO webserver is not available, you can install the prediction tool locally by following the instructions present in their GitHub repository at https://github.com/LBM-EPFL/PeSTo-Carbs.

7. A description of the format for the various restraint types supported by HADDOCK can be found in our Nature Protocol paper, Box 1. Information about various types of distance restraints in HADDOCK3 can also be found in our online haddock3 user manual pages.

8. Installing the latest version of HADDOCK3 by cloning its GitHub repository (section 3.1.2) will give you direct access to workflow examples, including protein-glycan ones.

9. In HADDOCK3, one can obtain the list of general parameters or module specific parameters, together with their description and default values using the `haddock3-cfg` command line interface.

    To obtain the list of general parameters, simply run the `haddock3-cfg` command

    ```
    haddock3-cfg
    ```

    To obtain the list of module specific parameters, run the `haddock3-cfg` command using the `-m` option followed by the module name (e.g. for `rigidbody`):

    ```
    haddock3-cfg -m rigidbody
    ```

10. The tuning the value of the `w_vdw` parameter to 1.0 (`w_vdw = 1.0`), instead of the default value of 0.1, is a recommended setting also for small molecule docking.

11. Adding a `caprieval` module after each module generating models in the workflow allows us to analyse the statistics of the generated models and their convergence. Placed after a selection module, it can pin-point an issue in the scoring function and therefore the ranking of the models, while placed after a refinement method, such as the flexible refinement one (`flexref`), it can show how the models (hopefully) improve.

12. The `caprieval` module supports the use of a reference file that can be provided. This is a special case, where the result of the docking is already known experimentally, and used to benchmark protocols. In a real world scenario, when the solution is not known and no reference is provided, the best scoring model is used as reference instead.

13. In the docking workflow, a clustering step is placed between the rigid-body docking and the flexible refinement stage. This protocol is specific to this protein-glycan workflow, as it allows to bring models potentially ranked low to the next stage, instead of selecting the best models from the rigid-body docking. This strategy is used as the HADDOCK scoring function is not perfect and therefore good models could be ranked too low to be forwarded to the next stage.

14. In HADDOCK, model generated and values reported can vary depending on the hardware on which you are running the simulations, and therefore reported values in this protocol may differ from the ones you will obtain.

15. This result is obtained using ideal 3.9Å ambiguous interaction restraints obtained from the reference structure, and it must be stressed that HADDOCK is sensitive to the quality of the input restraints. Therefore, using a less precise definition of interacting residues will often lead to poorer results.

16. Among the tools present in the HADDOCK ecosystem, the haddock-restraints (https://github.com/haddocking/haddock-restraints) is made to generate ambiguous and/or unambiguous interaction restraints. This tool can be installed locally as standalone but also has an online graphical interface available at https://wenmr.science.uu.nl/haddock-restraints.